# Feedback between population and evolutionary dynamics determines the fate of social microbial populations

Alvaro Sanchez and Jeff Gore

Department of Physics, Massachusetts Institute of Technology, 77 Massachusetts Ave, Cambridge, MA 02139

Correspondence should be addressed to:

Jeff Gore, Building  13-2008, 77 Massachusetts Avenue, Cambridge, MA 02139.

Email: gore@mit.edu, Phone number:  617-715-4251

Alvaro Sanchez, Building  13-2054, 77 Massachusetts Avenue, Cambridge, MA 02139.

Email:  alvaros@mit.edu

**ABSTRACT**

The evolutionary spread of cheater strategies can destabilize populations engaging in social cooperative behaviors, thus demonstrating that evolutionary changes can have profound implications for population dynamics. At the same time, the relative fitness of cooperative traits often depends upon population density, thus leading to the potential for bi-directional coupling between population density and the evolution of a cooperative trait. Despite the potential importance of these eco-evolutionary feedback loops in social species, they have not yet been demonstrated experimentally and their ecological implications are poorly understood. Here, we demonstrate the presence of a strong feedback loop between population dynamics and the evolutionary dynamics of a social microbial gene, SUC2, in laboratory yeast populations whose cooperative growth is mediated by the SUC2 gene. We directly visualize eco-evolutionary trajectories of hundreds of populations over 50-100 generations, allowing us to characterize the phase space describing the interplay of evolution and ecology in this system. Small populations collapse despite continual evolution towards increased cooperative allele frequencies; large populations with a sufficient number of cooperators "spiral" to a stable state of coexistence between cooperator and cheater strategies. The presence of cheaters does not significantly affect the equilibrium population density, but it does reduce the resilience of the population as well as its ability to adapt to a rapidly deteriorating environment. Our results demonstrate the potential ecological importance of coupling between evolutionary dynamics and the population dynamics of cooperatively growing organisms, particularly in microbes. Our study suggests that this interaction needs to be considered in order to explain intraspecific variability in cooperative behaviors, and also that this feedback between evolution and ecology can critically affect the demographic fate of those species that rely on cooperation for their survival.

**BLURB**

A new study finds that the evolution of social genes may be coupled with population dynamics, and dramatically affect ecological resilience, particularly in the face of rapidly deteriorating environments.

**INTRODUCTION**

Evolutionary changes in a species can strongly affect its environment over the timescales where speciation typically occurs. While this long-term effect of evolution on ecology has been long appreciated, it is typically assumed that evolutionary dynamics occurs over timescales that are too long to affect the dynamics of population size in the short term[1]. For this reason, most models of population biology ignore evolutionary changes in the different species, implicitly assuming a separation of timescales between population dynamics and evolutionary biology [2]. However, recent experiments suggest that changes in allele frequency can occur over timescales that are comparable to those typical of population dynamics [1,3–6]. Given this overlap in timescales, evolutionary dynamics and population dynamics may be coupled in what has been termed an eco-evolutionary feedback loop [1,3].

These eco-evolutionary feedback loops are predicted to be particularly strong in cooperatively growing species [7–11], which produce common goods and typically have larger fitness at large population densities than at low population densities [12–14]. Cooperative species can be challenged by the emergence of intraspecific "cheater" phenotypes, which take advantage of the common good produced by the community but do not contribute to its production. As a result, the cheaters may have higher fitness than cooperators and proliferate in the population at the expense of cooperators. The decline in cooperator numbers driven by evolutionary competition with the cheaters can have strong ecological consequences, as the ability of the population to produce the common good may be compromised [13]. These interactions have been predicted theoretically to yield an eco-evolutionary feedback between the allele frequency of a cooperative gene and the population size [7,8,10,11]. However, this bi-directional feedback has not been demonstrated experimentally, and the ecological consequences of such feedback are not known.

Microbes are remarkably social organisms [15], and are also amenable to laboratory experimentation. Very often, microbial cooperation results from the secretion of "public goods" to the media, such as quorum sensing molecules, extracellular enzymes, or the polymers that make up the fabric of biofilms. In some microbial species, population dynamics has been found to influence the evolution of cooperation via density-dependent selection [12,13,16,17]. In turn, it has also been found that, for some cooperatively growing species, the evolutionary competition between cheaters and cooperators can affect the growth of yeast [18,19] and bacterial [20] populations. Therefore, we reasoned that microbial ecosystems are likely candidates to display these predicted eco-evolutionary feedback loops.

In this paper, we characterized an eco-evolutionary feedback loop in a social microbial species. Given that the secretion of public goods in microbes is ubiquitous [15], these eco-evolutionary feedback loops may play crucial roles in microbial ecosystems. Our aim is to investigate whether eco-evolutionary feedbacks do indeed play a role on the evolutionary dynamics of cooperative traits, and what effect they play in the ecological properties of the populations where the evolution of cooperation is taking place.

## RESULTS

### Evolutionary dynamics of the SUC2 gene dramatically alters population dynamics

 To explore these eco-evolutionary feedback loops experimentally, we utilized the cooperative growth of budding yeast in the sugar sucrose. This cooperative growth is mediated by a single cooperative gene, SUC2, which codes for invertase, an enzyme that breaks down sucrose into glucose and fructose [13]. Invertase is secreted to the periplasmic space between the cell membrane and the cell wall [18]. As a result of this location outside the membrane, ~99% of the glucose and fructose produced by invertase diffuses away to be consumed by other cells in the population, while only the remaining 1% is directly captured by the cell that produced it [18]. This behavior leads to a cooperative transformation of the environment by the cells: at low population density, the cells are too dilute to effectively transform the sucrose environment into a glucose environment, so the cells grow slowly on what little glucose they retain following sucrose hydrolysis. At high population density, however, the cells are able to produce enough glucose for the population to grow rapidly (as found in ref [14] and in Figure S1). Because of this density-dependent cooperative growth, a minimal starting population size is needed to survive successive growth-dilution cycles on batch culture [14,18] (Figure 1A,B, Materials and Methods). In the absence of evolutionary dynamics (SUC2 gene frequency of 100%), we observe either rapid collapse or rapid approach to a stable population size.

The effect of SUC2 evolutionary dynamics on the population dynamics was assessed by growing mixed cultures of SUC2 carriers (cooperators), and a second strain with a SUC2 deletion (cheaters) [18]. Each strain was transformed with a fluorescent protein of different color, so cheaters and cooperators could be discriminated by flow cytometry (see Materials and Methods). Four cultures were inoculated with different initial SUC2 frequencies (from f=0.05 to f=0.5) and initial cell densities ranging from $N=10^3$ to $N=10^4$ cells/$\mu$L, and were then subject to a daily growth-dilution cycle (667x dilution factor) for five days. We found that the population dynamics are much more complicated than they were in the absence of evolutionary dynamics, with multiple cultures displaying seemingly erratic, non-monotonic changes in population size and in frequency of the SUC2 gene.

This experiment shows that evolutionary dynamics of the SUC2 gene, which is essential for cell growth under the conditions of the experiment, causes a dramatic change in the population dynamics. However, it is difficult to appreciate any specific patterns given the widely different and seemingly erratic behavior of both population and evolutionary dynamics when plotted separately. To gain insight into their relationship, we plot the trajectories followed by the different populations in an eco-evolutionary phase space formed by population density on one axis and the frequency of SUC2 on the other (Figure 1E). We find that these eco-evolutionary trajectories "spiral" in the density / frequency phase space, suggesting the presence of a coupling between population and evolutionary dynamics.

This feedback can be captured by a simple phenomenological model that naturally incorporates the coupling between evolutionary dynamics and population dynamics (See references [14,16] and Supporting Information). The model assumes that the growth rate of all cells in the population depends on the density of SUC2 carriers (cooperators) in the population. Below a certain threshold, cooperator cells grow at a slow rate on what little glucose they retain, while "cheaters" grow even more slowly. Above that threshold, both cooperators and cheaters grow at a fast rate drawing from the common pool of glucose, but cheaters grow faster as they do not have the metabolic burden of expressing the SUC2 gene. The growth of both phenotypes below and above the cooperator threshold is described by coupled logistic equations, to account for the fact that cooperators and cheaters compete for the glucose made by the cooperators, as well as other metabolites in the media (Figure 1F and Figure S1). This simple model predicts an eco-evolutionary phase space that is remarkably similar to our experimental measurements, with a separatrix dividing the phase space into two regions (Figure 2A). For population sizes larger than the separatrix, trajectories spiral to an eco-evolutionary equilibrium state characterized by co-existence between the cooperator and cheater phenotypes. For population sizes smaller than the separatrix, trajectories go extinct despite the fact that cooperators increase in frequency in the population (Figure 2A).

**Direct visualization of eco-evolutionary trajectories reveals the presence of a feedback loop between population and evolutionary dynamics of the SUC2 gene**

To test the phase-space mapping predicted by our model we scaled up the experiment and started sixty independent cultures, varying both the initial cell density and the initial frequency of the SUC2 gene in the population. Each of these cultures was subjected to daily growth-dilution cycles and both the cell density and frequency of the SUC2 gene were measured daily over the course of five days. We found a striking confirmation of the global eco-evolutionary feedback represented by spiral trajectories in the phase plane (Figure 2B). As predicted by the model, above the separatrix populations spiral to an equilibrium fixed point $d_{eq}$ (N=5.78±0.21×$10^4$ cells/$\mu$L, f=0.086±0.007; Mean±SE,N=3), while below the separatrix populations go extinct. In order to visualize this spiraling behavior close to equilibrium, we repeated the experiment by starting sixty mixed populations close to equilibrium, and followed them for eight days. The spiraling behavior was confirmed, as shown in Figure 2C. This experimentally observed behavior is consistent with the trajectories theoretically predicted by ecological public goods games [7,8,11].

**The evolutionary spread of cheaters does not cause early population collapse and does not significantly affect the productivity of the population**

The mapping of the eco-evolutionary space described above allows us to determine the fate of a cooperator population that is invaded by a cheater phenotype. A population of cooperators in equilibrium at $c_{eq}$ ($5.96 \pm 0.16 \times 10^4$ cells/$\mu$L, f=1.0; Mean$\pm$SE,N=3) that gets invaded by a $\Delta$SUC2 cheater mutant still falls to the right side of the separatrix (See Figure 3A, where we represent, in light gray arrows, the trajectories for all of the populations we measured). Therefore, rather than collapsing, the population will spiral to the new eco-evolutionary fixed point $d_{eq}$. Furthermore, the size of the population at equilibrium in $d_{eq}$ is very similar (smaller by less than 10%) to that in the pure cooperator population $c_{eq}$, indicating that the community can be supported by a relatively small fraction of cooperators.

**The evolutionary spread of cheaters decreases population resilience**

Given the modest deleterious effects caused by the spread of cheaters in the population, we wondered whether ecological properties might be affected by the presence of cheaters. We first noticed that while the population size in the eco-evolutionary equilibrium point $d_{eq}$ is very similar to the population size for a pure cooperator population $c_{eq}$, the distance between $d_{eq}$ and the separatrix ($X_d$ ; Figure 3A) is much smaller than the distance between the pure cooperator equilibrium $c_{eq}$ and the separatrix ($X_c$ ; Figure 3A). This suggests that the resilience of the population in eco-evolutionary equilibrium is less than for a population of pure SUC2 carriers in equilibrium. To test this prediction, we performed a one-day dilution shock on six equilibrium populations of either pure or mixed populations. All six pure cooperator populations survived the one-day shock of dilution by a factor of 32,000X (as compared to the normal dilution by 667X before and after the shock), whereas all six populations at equilibrium with cheaters went extinct (Figure 3B). The presence of cheaters in the population therefore reduces the resilience of the population, even if the productivity of the population is unchanged. We quantified the resilience of both pure cooperator and mixed populations by repeating this experiment for ten different disturbance strengths, and determined the fraction of populations that recovered from the shock (Figure 3C). This experiment confirmed that the resilience of a mixed population in eco-evolutionary equilibrium at $d_{eq}$ is about 5 times smaller than for pure cooperator populations, as we expected from visual inspection of the eco-evolutionary phase space.

**Rapid environmental deterioration leads to population collapse in the presence of cheater cells**

Given the importance of timescales to the presence of eco-evolutionary feedback, it is natural to also consider the effect of varying the rate of environmental change, particularly in the context of deteriorating environments. Our model predicts that mixed populations at eco-evolutionary equilibrium can survive slow but not sudden environmental deterioration (Figure 4A and Figure S2). In contrast, the survival of a population of cooperators is predicted to be independent of the rate of environmental deterioration (Figure 4A and Figure S2). A population initially growing in a benign environment (characterized by a low dilution factor), finds an eco-evolutionary equilibrium point $d_{eq,1}$ at low SUC2 frequency, at the right side of the separatrix. In a harsher environment (characterized by a higher dilution factor), the fraction of cooperators at the eco-evolutionary equilibrium point $d_{eq,2}$ is predicted to be larger, and the separatrix line moves to the right and curves up (see Figure S2). Thus, the eco-evolutionary fixed point for a benign environment $d_{eq,1}$ may fall below the separatrix line for a harsher environment. As a result, if the dilution factor suddenly switches from a low value (benign environment), to a high value (harsh environment), a population that was previously in the eco-evolutionary equilibrium point for the benign environment $d_{eq,1}$, finds itself below the new separatrix and out of the basin of attraction of the new equilibrium, and therefore goes extinct.

We tested this prediction by first allowing six populations of pure cooperators and six mixed populations to reach equilibrium in a benign environment (667x dilution). We then subjected them to either rapid environmental deterioration by switching suddenly to a harsh environment (1,739x dilution), or slow environmental deterioration by increasing the dilution factor in two steps. As predicted, all of the pure cooperator populations were able to survive both fast and slow environmental deterioration (blue lines, Figure 4B-C). In contrast, while all of the mixed populations were able to adapt to the slow deterioration (Figure 4B), only one out of six adapted to the fast deterioration (Figure 4C and Figure S3). A similar outcome was observed when the two-step slow environmental deterioration was replaced by a gradually deteriorating environment (Figure S4). We therefore find that our populations in eco-evolutionary equilibrium are more sensitive to rapid environmental deterioration than are the pure cooperator populations.

## DISCUSSION

Cooperation by secretion of common goods is widespread in microbes; from the polymers that form the matrix of biofilms to the exo-enzymes that degrade complex organic matter [21]. Understanding how these cooperative traits are maintained in populations is an essential problem of deep importance not only in evolutionary biology, but also in microbial ecology and systems biology [15,17,22–26]. An essential feature of the eco-evolutionary feedback in our system is the fact that cooperators have preferential access to the common good that they produce [16,18,27].

This preferential access creates the density-dependent selection that favors cooperators at low densities and cheaters at high densities, which is essential for the feedback loop. Indeed, recent modeling work [9] has suggested that limiting the diffusion of a common good may result in eco-evolutionary equilibrium between cooperators and cheaters, and even predicts oscillatory behavior similar to our experimental observations [9]. Our findings may therefore extend to other microbial systems exhibiting similar patterns of density-dependent growth resulting from preferential access to the common good.

The presence of density-dependent selection provides a clear causal effect between population dynamics and evolutionary dynamics [28]. In addition to population density [12,17], other ecological factors such as disturbance frequency [29], population dispersal [30,31], resource supply [32,33], spatial structuring of populations [10,18,34], the presence of mutualisms [35–37] or the presence of a competing species in the environment [16], or often play an important role in the evolution of cooperation. The effect that these and other ecological factors play on the evolution of cooperation is well understood [34,38]. However, the reverse process, i.e. the effect that the evolution of cooperative traits may have on the ecological properties of populations is not as well understood [39]. Previous studies had found that under some conditions, the evolutionary competition between cooperators and cheaters may have effects on the productivity of the population [19] or in its growth rate [18]. The experiments reported here indicate that this effect of evolution on population dynamics further feeds back into the evolutionary competition between cheaters and cooperators.

Understanding the effects of rapid evolution in ecological systems [3,6,40–44], and in particular the possible emergence of feedback loops between ecology and evolution, has recently attracted great interest in the ecological and evolutionary biology communities [1,3,5,6,36,38,40–45]. In spite of their expected importance (and even though the idea that evolution and population dynamics may be coupled dates back at least to the 1960's; see [41] and references therein), the exploration of eco-evolutionary feedback between population and evolutionary dynamics and their ecological and evolutionary consequences is still in its infancy. Some recent studies have found that eco-evolutionary feedbacks may affect other ecological parameters such as the phase and period of predator-prey oscillations [42]. Our study highlights the potential importance of the coupled interaction between evolutionary and population dynamics in growing microbial communities, and suggests that this interaction needs to be considered in order to explain intraspecific variability in cooperative behaviors, and the demographic fate of those species that rely on cooperation for their survival.

## MATERIALS AND METHODS

*Strains*

Strains JG300A (cooperators) and JG210C carrying a SUC2 deletion (cheaters) were employed[15]. JG300A was derived from BY4741 strain of *Saccharomyces cerevisiae* (mating type a, EUROSCARF). It has a wild-type SUC2 gene, and constitutively expresses YFP from the *ADH1* promoter (inserted using plasmid pRS401 with a *MET17* marker). It also has a mutated *HIS3* (*his3Δ1*). JG210C is a *SUC2* deletion strain (EUROSCARF Y02321, *SUC2::kanMX4*), and constitutively expresses dTomato from the *PGK1* promoter (inserted using plasmid pRS301 containing a *HIS3* marker).

*Culture conditions*

Cells were grown in synthetic media (YNB and CSM-his; Sunrise Science, CA) containing 2% sucrose, 0.001% glucose, and 8 μg/mL histidine. Cultures were grown in the 60 internal wells of a Falcon flat-bottom 96-well plate (BD Biosciences, CA), each containing 200μL of the culture. The plates were incubated at 30°C, shaking at 800rpm. The external wells were filled with 200μL of growth media. The plate was covered with parafilm. Cultures were grown for 23.5 hr, and then diluted into fresh growth media by a 667x dilution factor, unless otherwise noted. The diluted samples were placed on a new plate, and incubated again for 23.5 hr. These serial growth-dilution cycles were interrogated for several days. Note that since earlier studies were performed in conditions where population density at the beginning of each growth cycle was kept constant, this eco-evolutionary feedback had not been observed before[18].

*Measurement of cell density and cooperator frequency*

At the end of each growth period, the optical density at 620nm on each well was determined with a Thermo Scientific Multiskan FC microplate spectrophotometer. A 10μL sample of each well was then transferred to a new plate containing 190μL sterile Cellgro PBS buffer (Mediatech, VA). These were then scanned at a high-throughput BD LSRII-HTS analyzer. Flow cytometry was used to determine the correspondence between cell density and the optical density measured at the plate reader (see Figure S5), as well as to identify cheaters and cooperators by their fluorescence emission (see Figure S6).

**ACKNOWLEDGEMENTS**

The authors would like to thank Lei Dai, Daan Vorselen and Hasan Celiker for their help during the initial stages of this project, and also Andrea Velenich and all other members of the Gore lab for helpful discussions. We also wish to thank Andrew Chen for collecting the data in Figure S1A.

**Financial Disclosure**

This work was supported by grant FQEB #RFP-12-07, NIH grants NIH DP2 and R00 GM085279-02, as well as grant PHY-1055154 from the NSF. The authors also received support from the Pew Foundation and the Alfred P. Sloan Foundation.

**Author Contributions**

A.S. and J.G. designed the experiments and discussed the results. A.S. did the experiments and analyzed the data. Both authors wrote the paper.

**Competing Interests**

The authors declare no conflict of interest.

**Figures for Main Text**

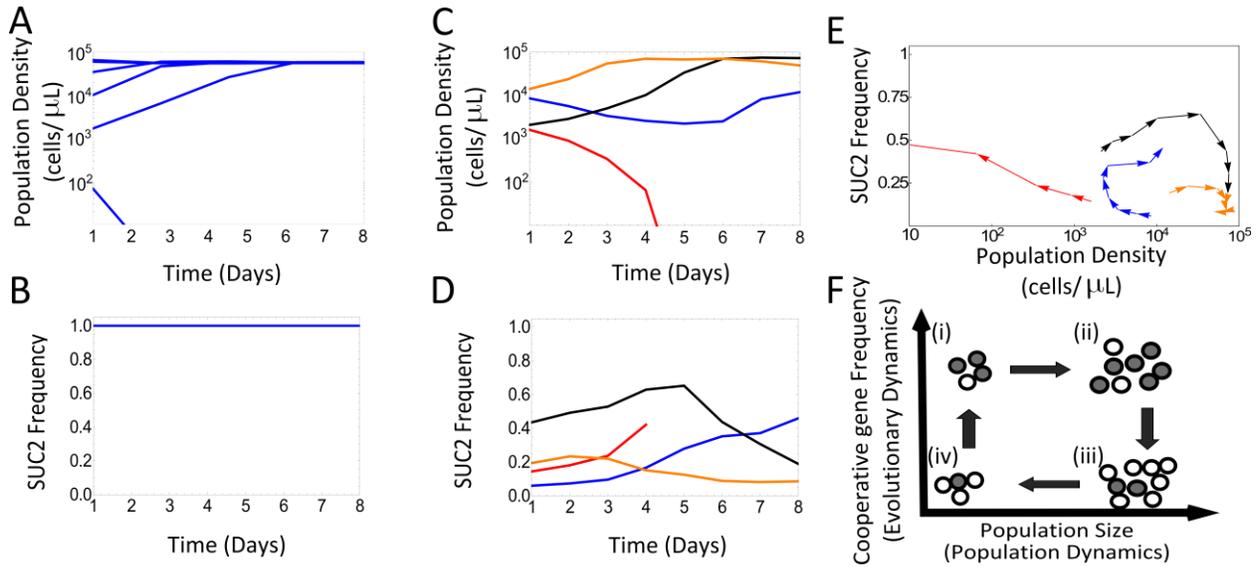

**Figure 1. Population dynamics in the presence and the absence of evolutionary dynamics**. Multi-day growth-dilution cycles demonstrate that evolutionary dynamics of a cooperative gene may dramatically affect population dynamics. (A-B) Yeast populations consisting exclusively of cooperator cells rapidly converge to an equilibrium population size in the absence of evolutionary dynamics.(C) Four different populations consisting of a mixture of SUC2 carriers and deletion mutants were subject to eight days of growth dilution cycles. Populations started at different population densities and SUC2 frequencies in the population. (D) Evolutionary dynamics for the same four populations as in (C) are represented by the same colors. Plots of the population and evolutionary dynamics show seemingly erratic, non-monotonic behavior. (E) By constructing an eco-evolutionary phase-space formed by the population size and the frequency of the SUC2 gene in the population, we find that the four populations in (C-D) follow well defined trajectories. Each trajectory corresponds to the evolutionary and population dynamics of the same color. (F), A simple conceptual model rationalizes the eco-evolutionary trajectories; Gray circles represent cooperators, white circles represent cheaters.

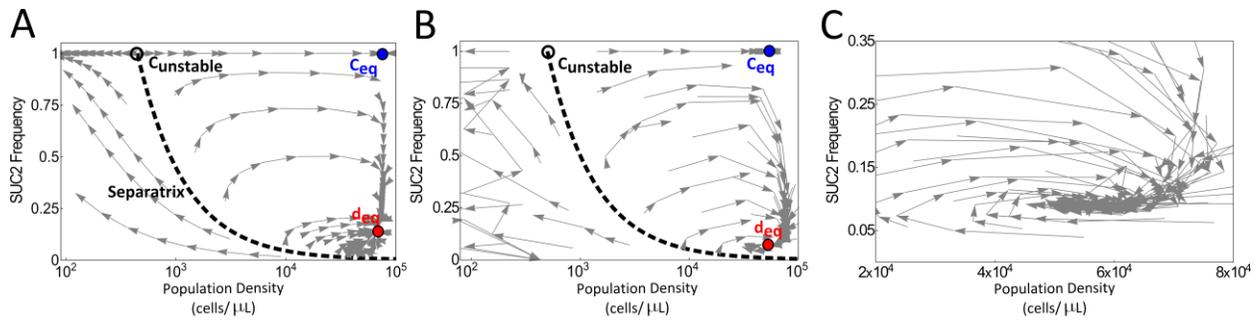

**Figure 2.** Visualization of eco-evolutionary trajectories. (A) Simulation of the eco-evolutionary growth model (see SI) over successive growth-dilution cycles. Gray arrows mark the day-to-day change in frequency of the SUC2 gene (f) and the population density (N). The eco-evolutionary phase space formed by N and f is divided in two regions by a separatrix line (black dashed). Above the separatrix, feedback between N and f results in trajectories spiraling toward an eco-evolutionary equilibrium point where cooperators and cheaters co-exist at deq (red dot). Below the separatrix populations go extinct despite the cooperators growing in frequency. In the absence of cheaters, the population dynamics have a stable fixed point at ceq (blue dot) and an unstable fixed point at cunstable (White circle). (B) Trajectories in the phase space for sixty cultures over five growth-dilution cycles. As predicted, a separatrix line divides the phase space in two regions: to the right trajectories spiral to an eco-evolutionary equilibrium and to the left trajectories lead to population collapse as cooperators increase in frequency. (C) A second set of sixty experimental populations were started in the vicinity of the co-existence equilibrium point deq and followed for eight days, further illustrating the spiraling behavior and thus the presence of a feedback loop.

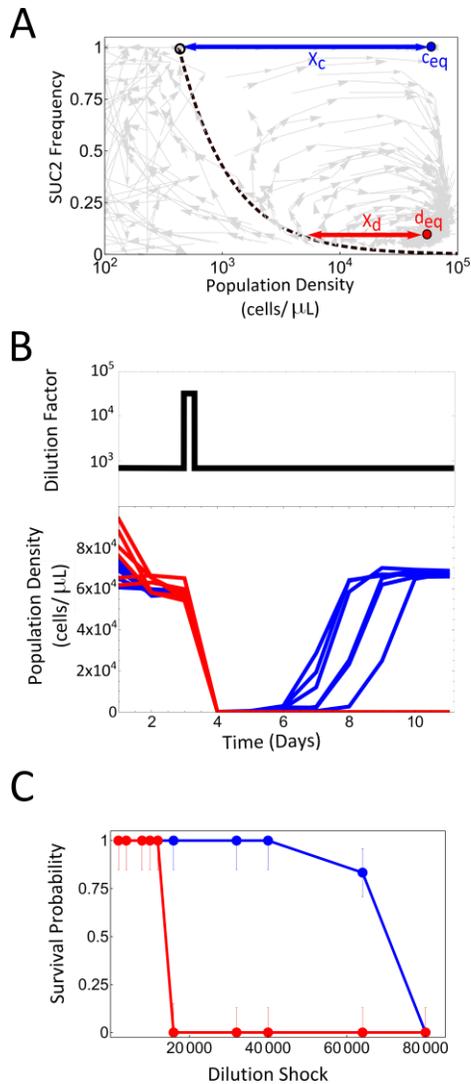

**Figure 3.** The presence of cheaters cells decreases the resilience of a population. (A) 180 eco-evolutionary trajectories corresponding to three different experiments are plotted in light gray. On top, we represent the population dynamics equilibrium point for pure cooperator cultures $c_{eq}$ (blue dot) and the eco-evolutionary equilibrium point $d_{eq}$ (red dot). The blue arrow marks the distance between $c_{eq}$ and the separatrix ($X_c$), and the red arrow marks the distance between $d_{eq}$ and the separatrix ($X_d$). (B) Populations were started near $c_{eq}$ (blue) or $d_{eq}$ (red) at a dilution factor of 667x. A large disturbance was applied on the third day of culture, by increasing the dilution factor to 32,000x for one day (top panel). Pure cooperator populations were able to recover, but the mixed cooperator/cheater populations in eco-evolutionary equilibrium went extinct. (C) Survival probability as a function of the strength of the perturbation (i.e. dilution shock). The presence of cheaters (red circles) decreases the population resilience, i.e. the maximum dilution shock that the population can withstand, relative to pure cooperator populations (blue circles). Error bars were estimated assuming binomial sampling (N=6), and represent a 68.27% confidence interval.

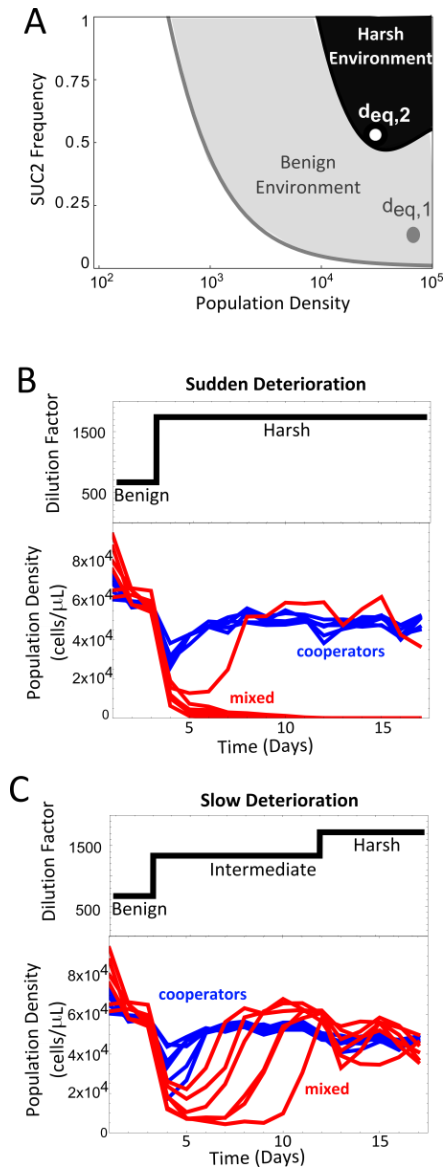

**Figure 4. The presence of cheaters makes a population unable to survive rapidly deteriorating environments.** (A), Schematic representation of the phase space for two different dilution factors, as obtained in simulations (see Figure S 2). For a benign environment (low dilution factor of 667x) the basin of attraction for the eco-evolutionary equilibrium point $d_{eq,1}$ is colored in light gray. For a harsh environment (characterized by a higher dilution factor of 1,739x) the basin of attraction for eco-evolutionary equilibrium point $d_{eq,2}$ is shaded in black. A population in the benign equilibrium at $d_{eq,1}$ that is suddenly switched to the harsh environment will go extinct, as it is out of the basin of attraction for $d_{eq,2}$. (B) This prediction was tested experimentally by bringing to equilibrium six pure cooperator populations and six mixed cooperator/cheater populations (all at a low dilution of 667x). The dilution factor was suddenly changed to 1,739x on day 3 (top panel). All six pure cooperator populations tested (lower panel, blue) were able to withstand the rapid deterioration. However, only one out of six mixed

populations (lower panel, red) that were originally near eco-evolutionary equilibrium in the benign environment (for a dilution factor of 667x) were able to survive the rapid environmental deterioration. (C) A slow environmental deterioration was applied by increasing the dilution factor from 667x to 1,739x in two steps (upper panel). In this case, all six mixed populations were able to survive the deterioration (as did all six pure cooperator populations).

## SUPPLEMENTARY FIGURES AND CAPTIONS

### Figure S1

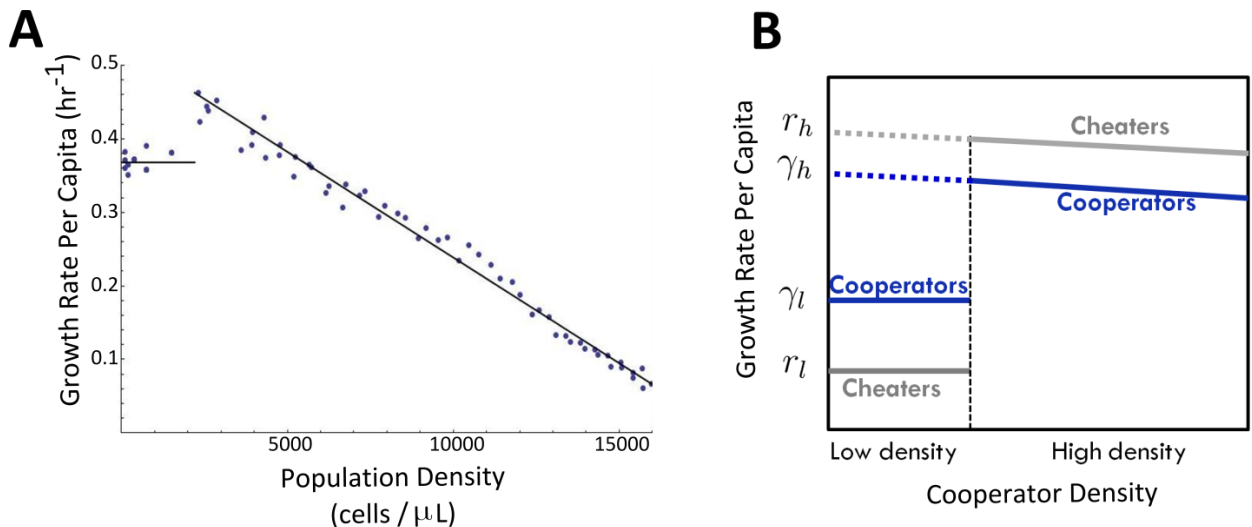

**Figure S1: Cooperator growth is well described by a two-phase logistic growth model.** A, Cultures of the cooperator strain were grown at 30C for 20hr in 96-well BD microplates in the same growth media as in all other experiments in this article. The plate was incubated in a Varioskan Flash plate reader, which allowed us to automatically measure the optical density ($OD_{620}$) of the cultures every 15min. Cultures were started at different initial cell densities, which allowed us to determine the growth rate as a function of density and distinguish two regimes. The growth rates at low and high densities were obtained from the raw data as previously described[14,18]. We plot here the growth rate per capita as a function of cell density (blue dots), and find that it is well fitted by the bi-phasic logistic model describe in equation SI-1 (black line). This indicates that the bi-phasic logistic growth model is a reasonable phenomenological model for our experiments. Note that the growth conditions differed substantially from our other experiments in the following: (i) The plates were not covered with parafilm, which may have resulted in different levels of oxygen in the sample, as well as increased evaporation; (ii) the plates were not shaken continuously, but only for a period of 2min immediately preceding OD measurement; and (iii) the environment of the plates was not an incubator, but at a plate reader, so that the temperature controls were presumably different. Therefore the quantitative parameters extracted from the fit to the growth curves, cannot be directly extrapolated to our experiments. B,

Schematic illustration of the bi-phasic Lotka-Volterra model of competition between cooperators and cheaters. The growth rate for cooperators and cheaters is represented as a function of cooperator density (note that this cartoon is a simplification, whose purpose is to develop intuition about the meaning of the different parameters). We wish to express our gratitude to Andrew Chen for collecting the data presented in A.

**Figure S2**

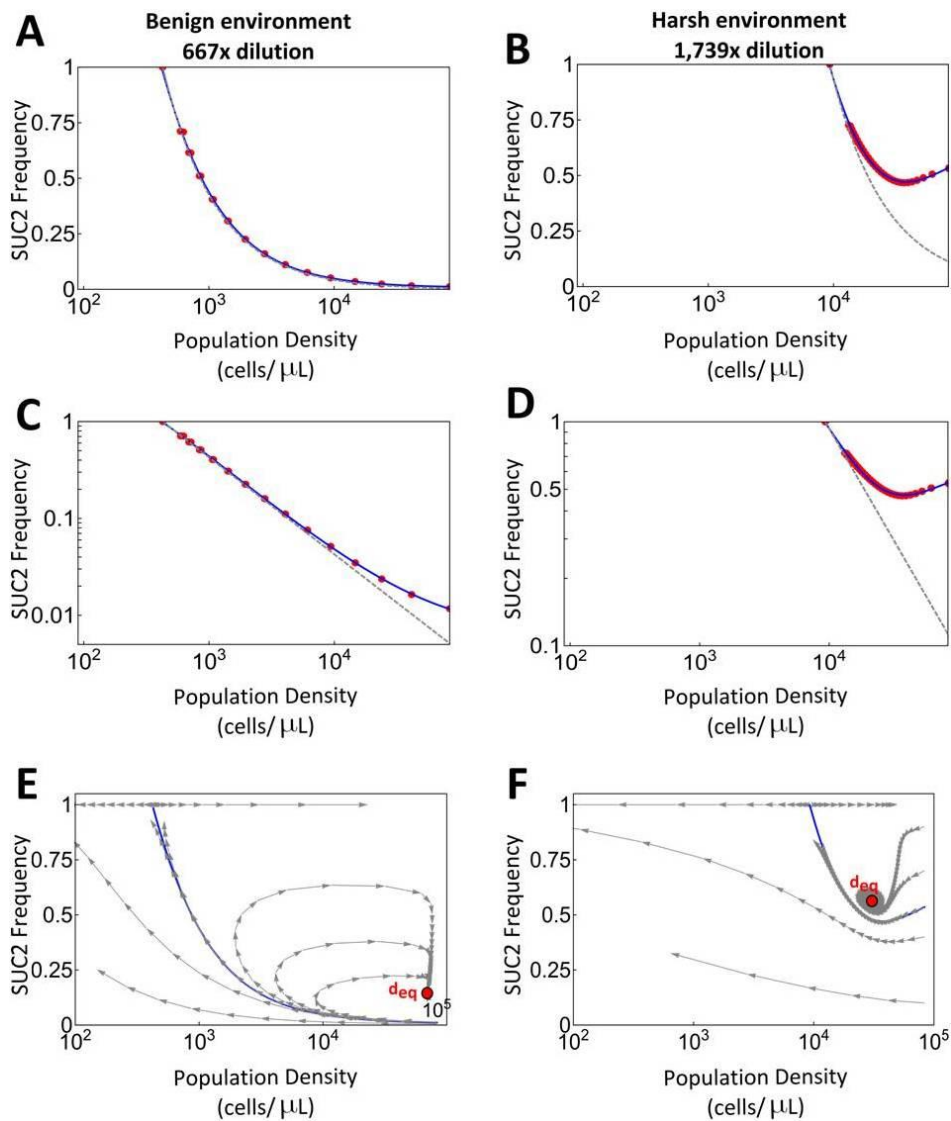

**Figure S2: Estimation of co-existence fixed points and their basin of attraction from the model.** In order to estimate the location of the separatrix (the line that marks the basin of attraction of the coexistence fixed points $d_{eq}$) at different dilution factors, we started the simulations at carrying capacity and multiple different values for the cooperator fraction. The separatrix locations for a low dilution factor (667x) (A,C) and a high dilution factor (1,739x) (B,D) are shown in linear and log-linear scales (red dots), together with a polynomial fit (blue lines) and the prediction of a null model ignoring eco-evolutionary feedback (gray dashed line), which corresponds to the curve in the eco-evolutionary space where the cooperator density is $N_c$ = $c_{unstable}$; i.e. the curve given by $f = c_{unstable}/N$. As one would expect, the separatrix follows closely the null model at low population densities (where Nc+Nd/K<<1 and the feedback is weak) and deviates from it at higher population densities (where Nc+Nd~K and feedback is stronger). Note that for a dilution factor of 667x, $c_{unstable}$=440 cells/$\mu$L, which is well below the carrying capacity of 83,341 cells/$\mu$L. Also, we note that the deviation from the null model is larger at the harsh environment (dilution factor 1,739x) than at the benign environment. This is because $c_{unstable}$ =9,330 for the harsh environment, which is much larger than for the benign environment, and already close to the carrying capacity. For this reason, the separatrix deviates sooner from the null model prediction. In E,F we simulate trajectories that were started near to the separatrix, on top of the the separatrix, or below the separatrix for both benign (E) and harsh (F) dilution factors (blue).

**Figure S3**

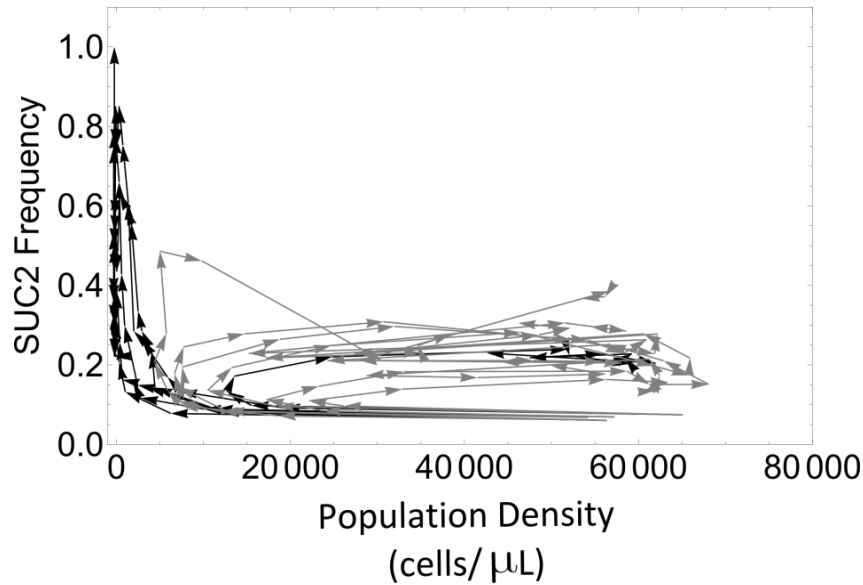

**Figure S3: Effect of fast and slow environmental deterioration on the eco-evolutionary phase space.** The data represented in figure 4 is projected into the eco-evolutionary phase space. Black and grays arrows represent the eco-evolutionary trajectories associated to figures 4B (rapid deterioration) and 4C (slow deterioration), respectively.

**Figure S4**

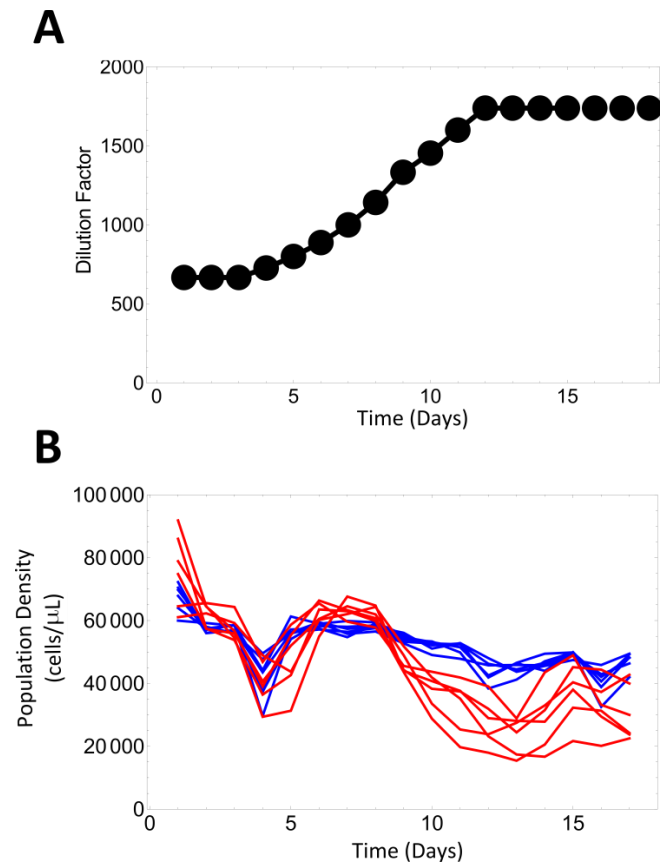

**Figure S4: Adaptation to gradual environmental deterioration.** The experiment in figure 4C was repeated but, rather than changing the environment in two steps, we slowly increased the dilution factor (A) from 667x to 1739x. B, All populations, both pure (blue) and mixed (red), survived the slow deterioration.

**Figure S5**

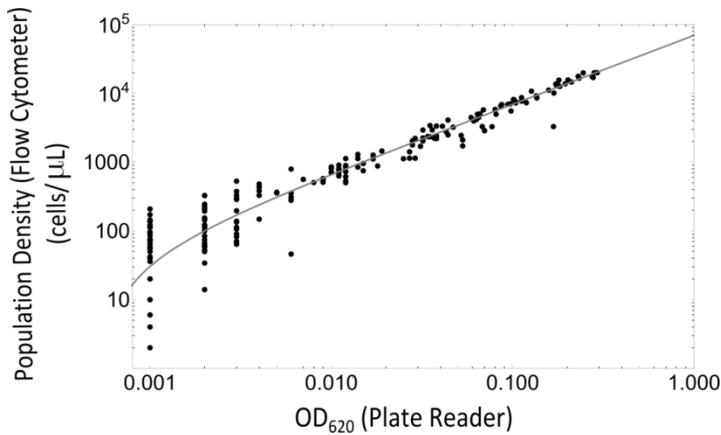

**Figure S5: Calibration flow cytometer – OD meter.** A calibration is performed to quantify the relationship between cell density (as determined by flow cytometer analysis, which allows us to count the number of cells in 10 uL cultures), and optical density ($OD_{620}$). The relationship between the two is linear; we obtain a reasonable fit to the line y=14.52+69561 x (solid gray line)

**Figure S6**

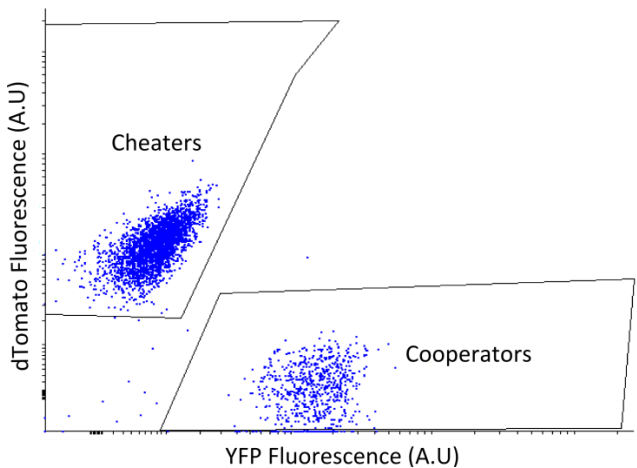

**Figure S6: Separation of Cheaters and Cooperators by the flow cytometer.** Typical data corresponding to flow cytometry analysis of mixed cultures suspended on PBS media. Cooperators and cheaters form two distinct populations in the space formed by yellow and red fluorescence emission; cooperators express YFP constitutively, and therefore have strong emission in the yellow, but low emission in the red; cheaters express a red protein, dTomato, and therefore have strong emission in the red, but low emission in the yellow. Individual cells could thus be identified as one or the other by virtue of their different spectral fluorescence emission.